06-19-06

Free-energy distribution of binary protein-protein binding suggests cross-species interactome differences.

Yi Y. Shi[1], Gerald A. Miller[2], Hong Qian[1], Karol Bomsztyk[3]

Departments of Applied Mathematics[1], Physics[2] and Medicine[3], University of Washington, Seattle, WA 98195

Address for correspondence:

 Karol Bomsztyk

Box  358050

University of Washington,

Seattle, WA 98109 USA

Tel. (206) 616-7949

Fax. (206) 616-8591

 e-mail:  karolb@u.washington.edu

**Abstract**

Major advances in large-scale yeast two hybrid (Y2H) screening have provided a global view of binary protein-protein interactions across species as dissimilar as human, yeast, and bacteria. Remarkably, these analyses have revealed that all species studied have a degree distribution of protein-protein binding that is approximately scale-free (varies as a power law) even though their evolutionary divergence times differ by billions of years. The universal power-law shows only the surface of the rich information harbored by these high-throughput data. We develop a detailed mathematical model of the protein-protein interaction network based on association free energy, the biochemical quantity that determines protein-protein interaction strength. This model reproduces the degree distribution of all of the large-scale Y2H data sets available and allows us to extract the distribution of free energy, the likelihood that a pair of proteins of a given species will bind. We find that across-species interactomes have significant differences that reflect the strengths of the protein-protein interaction. Our results identify a global evolutionary shift: more evolved organisms have weaker binary protein-protein binding. This result is consistent with the evolution of increased protein unfoldedness and challenges the dogma that only specific protein-protein interactions can be biologically functional..

**Introduction**

Gaining a global view of protein-protein interaction (PPI) networks gives a new perspective to the understanding of all biological organisms (1-3). Advances in yeast two-hybrid (Y2H) interaction have provided high-throughput readouts that generate maps of PPI networks in several organisms including man (4-11). These large-scale interactomes revealed an approximately scale-free-like topology that is shared by each studied species. This means that in all organisms most proteins have one or two partners, but a few (so called hubs), have many partners. Thus the probability $p(k)$ that a protein interacts with $k$ others follows an approximate power-law distribution: $p(k) \propto 1/k^{\gamma}$. This conserved cross-species PPI property is not surprising because networks with power-law distributions are ubiquitous appearing in systems as diverse as the internet, the citation index and societies (12-14).

The discovery of a common topology of diverse systems whose functions are so strikingly different initiated the search for universal models to explain scale-free networks (13, 15). The concept of preferential attachment, in which in growing networks new vertices link preferentially to older nodes that are already highly connected (13), is very popular. Analysis of species separated by billions of years of evolution showed that this mechanism could also be involved in evolutionarily expanding protein-protein networks (16). Another newer idea of intrinsic fitness, in which two nodes are connected when the link is mutually beneficial, was proposed to explain scale-free networks (15). This class of models shows that intrinsic fitness is an essential property that underlies the power-law distribution and also allows the prediction and measurement of other properties. In particular, an exponential distribution of the fitness leads to a power law degree distribution of a network.

Protein-protein binding is determined by free energy of association as well as the concentrations of participating molecules (17). The biochemical manifestation of intrinsic fitness for protein-protein binding is that each protein has an inherent propensity for association. This idea opposes the view that protein-protein interactions are determined solely by a "lock-and-key" mechanism involving complementarity. This paper uses the properties of protein-protein interaction networks to quantitatively explore the unorthodox view of protein interaction promiscuity.

The Y2H method reports binary results for protein-protein binding under a controlled setting (18). We assume that a Y2H measurement is an efficient way to measure a binary protein-protein interaction, just as one can do for a pair of proteins in a test tube. Thus, the association reaction of two proteins, say $A$ and $B$, is determined by the free energy difference $\Delta G^o$ (19) between the state $A+B$ and the $AB$ final state (Fig.1). The large-scale Y2H data sets report the presence of an $AB$ complex.

We discuss this in more detail. In Y2H screens, two fusion proteins are generated: one protein, is constructed to have a DNA binding domain attached to its N-terminus, and its potential binding partner, is fused to a transcriptional activation domain (18). Binding of the two proteins will form an intact and functional transcriptional activator. This newly formed transcriptional activator complex will then transcribe a reporter gene whose protein product can be assayed. Thus, the presence of the reporter gene product generated is a measure of the association between two proteins. The probability of two proteins forming a complex is determined by their association constant, $K_a$, which is in turn related to the free energy measured in unit of $RT$.

In the large-scale Y2H screens concentrations of all expressed hybrid proteins are expected to be approximately the same. Hence, the factor of protein concentrations, which surely plays a role in vivo, is negated in the binary interactions measured in the Y2H system. This is in contrast to PPI measurements using mass spectroscopy that depend on the native protein concentrations in cells (20). In this report, we focus on the data derived from Y2H screens and hence on the strength of protein-protein binding alone. The physical origin of protein-protein interaction strength can be complex: hydrogen bonding, van der Waals, hydrophobic, and electrostatic interactions all play a role. However, a thermodynamic model can be developed irrespective of the nature of the free energy difference, $\Delta G^o$. In this study we developed a quantitative model using both exact simulation technique and semi-analytic approximation to test whether the current large-scale Y2H binding data sets obtained for multiple species can be interpreted in terms of a distribution of $\Delta G^o$ of an organism; and furthermore if distributions of free energy of interactions differ across species.

**Results and Discussion**

The overall strategy used to derive organism's free energy distribution of binary protein-protein interaction is illustrated in Fig. 1 and detailed in Method section.

Large-scale Y2H screens of protein interactions have been completed for several organisms including the bacteria *H. pylori* (8), the malaria parasite *P.falciparum* (6), yeast *S. cerevisiae* (5, 11), worm *C. elegans* (7), fruit fly *D. melanogaster* (4) and human (9, 10). For all the organisms examined the distribution of Y2H protein-protein interactions approximately follows a power law (4-6, 8-11) form. We sidestep the issue of whether the PPI topology is exactly scale-free. Instead we apply our model, which uses free energy as the basis of the of protein-protein interaction in a thermodynamic approach to understanding protein-protein interactions, to describe the Y2H data available for the different species. In this approach the power-law behavior of networks is derived from an exponential distribution giving the probability of variations in the free energy contributed by a protein. Our starting point is the protein-protein interaction and the additivity principle (21). Simulation and semi-analytic strategies were used in a complementary fashion (Methods).

The computer-simulated fit and semi-analytically derived curves superimposed on the Y2H data for each organism are shown in Fig. 2. Modeling the data obtained from the large-scale Y2H screens (4-6, 8-11) using our approach yielded two parameters, $\lambda$ and $\mu$, for each species (Table 1). The high-throughput Y2H maps represent a partial sample of the interactomes. Questions have been raised about the accuracy of inferring a complete PPI topology from only a partial sample (22). In our model, the nearly identical $\lambda$ and $\mu$ parameters derived for the two independent human high-throughput Y2H screens not only provide independent validation of those large-scale data sets, but also support the current model. The analytically-derived curves for the different species data sets reveal that the degree distribution resembles but does not strictly follow a power law. Importantly, unlike the prior analysis (10), the current model reveals species differences in the parameters that control the degree distribution. The value of $\lambda$ ranges from 0.64 to 1.53, and is closely related to the slope of the curves representing $p(k)$ (Fig. 3A). The value of $\lambda$ reflects the tightness of the fluctuations of the free energy difference, with a smaller value indicating increased fluctuation of the ability to interact. There is no obvious correlation between $\lambda$ and divergence times among these organisms. The parameter $\mu$ is closely related to the height of the curves representing $p(k)$ (Fig. 3B). This parameter is a measure of the average association free energy difference and therefore can be regarded as an indicator of the average strength of all the binary protein-protein interactions of an organism. The value of $\mu$ differs across species and unlike $\lambda$ is positively correlated with divergence times (Fig. 4). This means that $\mu$ is lowest for *H.pylori* and progressively increases with increased evolutionary time. In other words, the average strength of binary protein-protein interactions is strongest in the least complex organism.

Recently Deeds et.al. proposed a physical model for protein-protein interactions based on number of exposed hydrophobic residues that similarly recapitulates power law distribution in yeast (23). Their model is a specific example of the "intrinsic fitness model" advanced by Caldarelli et.al (15), with the fitness number of each node having Gaussian distribution and the probability of interaction $p(g,g')$ taken as a step function. Our model and their model each obtain degree distributions that are approximately scale free. While their approach based on the Gaussian distribution only holds for a small range of parameters, our model based on exponential distribution yields a nearly scale-free distribution for almost any set of parameters $\lambda$ and $\mu$. We have not been able to use their model to reproduce the measured degree distributions of all the species (not shown).

Our statistical model, as well as those of others (23), under-estimates the number of proteins with single partners. This is seen in Fig. 2 by comparing the empirical and predicted values of $p(k)$ at $k=1$. There are many classical biochemistry "lock-and-key" protein-protein interactions that involve highly specific pairs of proteins. Such interactions are not included in our statistical model that assumes additivity of free energy. Hence the experimental observation of $p(1)$ is always much greater than the predicted value. In fact, the traditional paradigm for structural-based protein recognition always emphasizes the complementarity between interactive pairs. This has led to the important concept of specificity in biochemistry. The intrinsic fitness-based model illustrates that "non-specific" interactions where one protein can have multiple partners play an important role in the large-scale protein-protein interaction. And more importantly, our result shows that such interactions can be biologically functional.

We can quantify the strong and weak interactions by computing the probability $h(\Delta g)$ for the association energy of a pair of proteins, $\Delta G^o / RT$, to take the specific value $\Delta g$. A straightforward computation yields the result

$$h(\Delta g) = \lambda^2 (\mu + 2/\lambda - \Delta g) e^{-\lambda(\mu + 2/\lambda - \Delta g)} \qquad (1)$$

The distribution $h(\Delta g)$ peaks when $\Delta g = \mu + \frac{1}{\lambda}$. The mean value of the distribution is $\mu$. The parameters $\lambda$ and $\mu$ for each species were used to generate the different energy distributions, $h(\Delta g)$ shown in Fig. 5.

The cross-species comparison of the free energy distributions shown in Fig. 5 reveals a progressive left-to-right shift of free-energy distribution with evolutionary time. This shifts towards weaker interactions mirrors changes in $\mu$ (Fig.4). There was a disproportionably larger difference between human and fly/worm free-energy distribution with respect to divergence time than the differences between fly/worm and the unicellular organisms. *Plasmodium* protein complexes network has diverged from those of yeast, fly and worm (24). Yet, we find that the free-energy distribution for this malaria pathogen is similar to these other early organisms (Fig.5).

What could be the evolutionary changes needed to account for the weaker interactions that seem to typify the human interactome compared to those of the lower organisms? Comparative genomic analysis reveals dramatic differences in the human proteome compared to lower metaozoans such as the fly or the nematode (25).

Disordered protein regions are common, particularly in regulatory factors (26). These domains can bind a diversity of protein partners (26, 27). It has recently been recognized that there is an increased trend towards protein unfoldedness from lower to highly complexed organisms (28). The unstructured protein domains are often modified post-translationally. These unfolded domains also permit multilateral binding and complex protein-protein interactions required by highly evolved organisms. This evolutionary change expands a protein's repertoire of partners and is a way for factors to assume new functions. The interactions involving disordered proteins are intrinsically weak. The ability to more readily dissociate a complex is considered an important attribute because it allows protein-protein interactions to be regulated by covalent modification and by other molecules.

Comparative genomic analysis also reveals other significant proteome changes that evolved in more complex organisms (25). For example, Src homology 2 (SH2) and Src homology 3 (SH3) bearing proteins are some of the most frequently represented families of factors in man, but their frequency in earlier species is orders of magnitude lower (25). The SH3 and SH2 interaction typically exhibit lower affinities, and are also highly regulated.

These evolutionary changes are embodied in the hnRNP K. This protein contains three structured RNA-binding KH domains that are well conserved in fly, nematode and yeast. KH domains are also found in bacteria (27). Mammalian K protein contains a large disordered KI region that contains several SH2- and SH3-binding sites that are absent in fly and worm. The KI region mediates association with many protein partners, interactions that are highly regulated by phosphorylation (29-31). In vitro, many K protein binary interactions are weak (29-31). Yet, within cells K protein is a component of many and functionally diverse complexes (27, 32). These observations may reflect multilateral molecular cooperativity of binding that is amiable to regulation by intra- and extracellular signals. K protein ability to interact in a regulated fashion with a diversity of other factors and nucleic acids explains its involvement in multiple processes that compose gene expression (27, 33). There are many mammalian proteins that exhibit similar properties (26).

Sequencing of many genomes revealed that the number of protein coding genes is surprisingly similar for organisms as disparate as human, fly and even yeast (25). Yet, the differences in the complexity of these organisms are immense. The large-scale Y2H screens across species provide opportunities to gain global views of the interactomes and their evolutionary trends. Our free-energy model of protein-protein interactions identifies for the first time a global evolutionary tendency towards weaker binary protein-protein Y2H interactions. The result of this analysis is consistent with the notion that in high complexity organisms disordered regions assumed a greater role (28). These weaker interactions became more important for human protein-protein interaction networks than for the networks of lower organisms, as viewed by Y2H screens. The evolution of weaker interaction, which is more easily modulated, provides new insight how cellular complexity could have evolved while maintaining genomic simplicity.

Undoubtedly, the future will bring many more large-scale Y2H studies. The model developed here should be useful for following interactome changes that evolved between more closely related organisms, and also for studying differences between the free-energy distributions of diverse tissues. In this regard it would be particularly interesting to compare the Y2H global view of the brain interactome to that of less complex organs. Comparing the free energy distribution of protein-protein interactions in normal and malignant tissues could also be very fruitful.

**Methods**

There are thousands of proteins in a typical protein-protein interaction network and millions of possible binary interactions. Therefore a statistical treatment of protein-protein interaction networks, based on the concept of free energy of association: $\Delta G^o = -RT \ln K_a$, where $K_a$ is the association constant, is used to derive the degree distribution of Y2H binary protein interactions for an organism.

*Derivation of the model*

Our basic ideas and assumptions are as follows. There is a mean association free energy among all the protein pairs in an organism: $\langle \Delta G^o \rangle$. For a particular pair of proteins, say $A$ and $B$, their association free energy deviates from the $\langle \Delta G^o \rangle$, and the deviation is contributed by both proteins $A$ and $B$ additively:

$$\Delta G^o_{AB} = \langle \Delta G^o \rangle - RT(g_A + g_B), \qquad (2)$$

where the $g_A$ and $g_B$ represent the fluctuations of the values of the free energy difference, measured in *RT* units, due to the respective contributions of protein *A* and *B*. We assume additivity following the theoretical work of (34) and (23) and molecular studies (35, 36). For general discussion of additivity principles in biochemistry, see (21). The empirical support for additivity assumption is also provided by the scale-free nature of Y2H protein-protein interaction networks (4-11). The scale-free phenomenon suggests that if *AB* has strong interaction, then *AC* is likely to have strong association. Conversely, if *XY* is a weak complex, then *XZ* is more likely to be weak. The physical basis for this intrinsic property of proteins to interact with others remains to be better defined. However, there is a correlation between the number of interactions by a given protein and the fraction of hydrophobic amino acids on its surface (23), suggesting one potential mechanism. Conventionally one would assume that $g_A$ and $g_B$ are Gaussian-distributed with zero mean. But it has been shown that the Gaussian distribution is inconsistent with networks exhibiting power-law topology (15). Rather, Caldarelli et al. have shown that the robust power-law topology essentially dictates the $g_A$ and $g_B$ to be exponentially distributed (15), thus it is asymmetric. The exponential distribution leading to power law behavior is also seen in the kinetic study of protein folding (37, 38). Therefore, we take the mathematical expression for the distribution of both $g_A$ and $g_B$, to be

$$\rho(g) = C \exp(-\lambda g), \quad \lambda \geq 0, \quad -1 \leq \lambda g \leq +\infty \qquad (3)$$

where *C* is a normalization factor whose value can be determined to be $\lambda/e$. The distribution of Eq. (3) has a mean of zero and standard deviation of $1/\lambda$. In summary the Y2H PIN power law topology (4-11) suggests using an exponential distribution (15) to define organism's free energy distribution of binary protein-protein interactions; the actual numerical value of the power seems to be related to the fluctuations of the protein-protein interaction.

We now ask, for a given protein $A$, what is the probability of it being associated with a protein $B$. This is a standard question of bimolecular association, and the probability is given by

$$p(\Delta G^o_{AB}) = \frac{K_{a,AB}[B]}{1+K_{a,AB}[B]} \qquad (4)$$

where $K_{a,AB}$ is the association constant between $A$ and $B$, $[B]$ is the concentration of molecule $B$. We assume that in all the Y2H experiments, the concentrations of the expressed hybrid proteins are essentially the same. Then Eq. (4) can be simplified into

$$p(\Delta G^o_{AB}) = \frac{e^{-\frac{\Delta G^o_{AB}}{RT}}[B]}{1+e^{-\frac{\Delta G^o_{AB}}{RT}}[B]} = \frac{e^{g_A+g_B-\mu}}{1+e^{g_A+g_B-\mu}}$$

where $g_A$ and $g_B$ are exponentially distributed according to Eq. (3), and the parameter $\mu = \frac{\langle \Delta G^o \rangle}{RT} - \ln[B]$ contains information on the average binding strengths of all the binary protein-protein interaction of a given organism. Its value is expected to be different for different species. Since we assume that all the Y2H measurements essentially have the same $[B]$, lower the mean association energy, greater the association constant, smaller the value of $\mu$. The $[B]$ in Eq. (4), the concentration of all the protein $B$ with its binding site for $A$ being free and independent of other binding sites, is a function of the concentrations of the other proteins that compete for the $A$ binding site of $B$. We do not take this effect into consideration in the present model.

In the remainder of this paper, we shall denote $\frac{e^{g_A+g_B-\mu}}{1+e^{g_A+g_B-\mu}} = p(g_A, g_B)$. It is graphically convenient to use a simpler notation in which the g-values of the protein-pair $A$, $B$ are denoted by $g$ and $g'$. Then for two interacting proteins with g-values $g$ and $g'$, the interaction probability $p(g, g')$ of Eq. (5) is determined solely by the quantity $g + g' - \mu$, with a positive value indicating a significant interaction probability.

$$p(g, g') = \frac{e^{g+g'-\mu}}{1+e^{g+g'-\mu}} \qquad (5)$$

Eq. (5) gives the probability of a protein $A$ forming a complex with another protein $A'$ leading to functional transcriptional activator. The relation between this probability and eventual expression of reporter gene is the complex transcription activation process. This is the least understood step of the Y2H measurement. There are essentially two types of transcriptional activation responses: graded and all-or-none (39-41). The latter leads to a step function as employed in (23). Reporter gene systems have revealed that at a single cell level expression is either maximal or not expressed at all, but the probabilities of expression are a function of the amount of transcriptional activators. This leads to graded responses in a cell population. We have adopted such a graded stochastic response in our simulations. A more detailed analysis and comparison of this aspect of Y2H measurements is in progress (manuscript in preparation)

*Computer simulation*

The use of random sampling techniques is appropriate for any system that can be described statistically, so we simulate the protein interaction network using the parameters $\lambda$, $\mu$ and $N$. The technique is to generate an $N \times N$ matrix with each element representing a chosen pair of proteins. A matrix element is 1 if the pair interacts or 0 if not. The first step in calculating the matrix element is to assign each protein a "g-value" according to the exponential probability distribution $\rho(g)$, Eq. (3). The second step is to calculate the probability of interaction $p(g, g')$ from Eq. (5). Increasing the value of $\mu$ decreases the value of $p(g, g')$ and therefore decreases the interaction probability, thus $\mu$ can be taken as an indicator of the strength of the interaction. Then a number $q$ ($0 \leq q \leq 1$) is generated randomly. If $p \geq q$ we say that there is an interaction between the two proteins and the matrix element corresponding to these two proteins is set to be unity. Otherwise, this element is set to zero. This procedure is repeated for all of the $\frac{N(N-1)}{2}$ pairs of proteins in the network. The sum of the number of ones in each row of the matrix represents the number of partners (or degree) of the chosen protein. Tabulating the degree of each protein allows us to determine $p(k)$ the probability that a protein has $k$ partners. Carrying out this procedure five times is sufficient to achieve a stable degree distribution.

*Semi-analytic approach*

We develop an average probability approximation to the exact formulation of (34). One of the consequences of the model represented by Eq. (5) is that the probability distribution of interaction free energy between protein *A* and all the other proteins is in fact different for different proteins. However, if we neglect this difference, and are only interested in the average distribution of interaction free energy between two proteins, a simple expression for *p(k)*, the degree distribution, can be derived. This is done by approximating the $p(g, g')$ by its average:

$$\bar{p}(g) \equiv \int_{-\frac{1}{\lambda}}^{\infty} \rho(g') p(g, g') dg' . \qquad (6)$$

Using the distribution given in Eq. (6), we build a network by assuming that a single protein with given a g-value binds another protein with a probability $\bar{p}(g)$. For a number of proteins, *N* (including itself), the probability of having *k* actually bind is given by a binomial distribution

$$p(k) = \int_{-\frac{1}{\lambda}}^{\infty} \rho(g) \binom{N}{k} (1-\bar{p}(g))^{N-k} \bar{p}(g)^k dg . \qquad (7)$$

Eqs. (6) and (7) give essentially the same degree distribution as the simulation procedure discussed above. Furthermore, if one replaces the probability of interaction of Eq. (5) by a step function, this model reduces to the intrinsic fitness models of refs. (15, 34). In that case the approximate Eqs. (6) and (7) give the same results as an exact treatment.

*Y2H Data Fitting*

The parameters $\lambda$, $\mu$ were varied so as to minimize the chi-squared parameter defined to minimize the difference between the logarithms of the theory $p(k)$ of Eq. (7) and the experimentally measured $p^{\exp}(k)$:

$$\chi^2 \equiv \sum_k \left( \frac{\log p(k)}{\log p^{\exp}(k)} - 1 \right)^2 . \tag{8}$$

The sum is over those values of $k$ for which $p(k) \neq 0$. The parameters from these fits were used in the simulation.

**Acknowledgement**

We thank Lynn Amon, Oleg Denisenko, Jay Hesselberth, Stan Fields, Tom Milac and Ram Samudrala for their input on this paper. This work was supported by NIH GM45134 and DK45978 (KB).

# References

1. Cusick, M., Klitgord, N., Vidal, M. & Hill, D. E. (2005) *Hum Mol Genet*.

2. Deane, C. M., Salwinski, L., Xenarios, I. & Eisenberg, D. (2002) *Mol Cell Proteomics* **1,** 349-56.

3. Salwinski, L., Miller, C. S., Smith, A. J., Pettit, F. K., Bowie, J. U. & Eisenberg, D. (2004) *Nucleic Acids Res* **32,** D449-51.

4. Giot, L., Bader, J. S., Brouwer, C., Chaudhuri, A., Kuang, B., Li, Y., Hao, Y. L., Ooi, C. E., Godwin, B., Vitols, E., Vijayadamodar, G., Pochart, P., Machineni, H., Welsh, M., Kong, Y., Zerhusen, B., Malcolm, R., Varrone, Z., Collis, A., Minto, M., Burgess, S., McDaniel, L., Stimpson, E., Spriggs, F., Williams, J., Neurath, K., Ioime, N., Agee, M., Voss, E., Furtak, K., Renzulli, R., Aanensen, N., Carrolla, S., Bickelhaupt, E., Lazovatsky, Y., DaSilva, A., Zhong, J., Stanyon, C. A., Finley, R. L., Jr., White, K. P., Braverman, M., Jarvie, T., Gold, S., Leach, M., Knight, J., Shimkets, R. A., McKenna, M. P., Chant, J. & Rothberg, J. M. (2003) *Science* **302,** 1727-36.

5. Ito, T., Chiba, T., Ozawa, R., Yoshida, M., Hattori, M. & Sakaki, Y. (2001) *Proc Natl Acad Sci U S A* **98,** 4569-74.

6. LaCount, D. J., Vignali, M., Chettier, R., Phansalkar, A., Bell, R., Hesselberth, J. R., Schoenfeld, L. W., Ota, I., Sahasrabudhe, S., Kurschner, C., Fields, S. & Hughes, R. E. (2005) *Nature* **438,** 103-7.

7.  Li, S., Armstrong, C. M., Bertin, N., Ge, H., Milstein, S., Boxem, M., Vidalain, P. O., Han, J. D., Chesneau, A., Hao, T., Goldberg, D. S., Li, N., Martinez, M., Rual, J. F., Lamesch, P., Xu, L., Tewari, M., Wong, S. L., Zhang, L. V., Berriz, G. F., Jacotot, L., Vaglio, P., Reboul, J., Hirozane-Kishikawa, T., Li, Q., Gabel, H. W., Elewa, A., Baumgartner, B., Rose, D. J., Yu, H., Bosak, S., Sequerra, R., Fraser, A., Mango, S. E., Saxton, W. M., Strome, S., Van Den Heuvel, S., Piano, F., Vandenhaute, J., Sardet, C., Gerstein, M., Doucette-Stamm, L., Gunsalus, K. C., Harper, J. W., Cusick, M. E., Roth, F. P., Hill, D. E. & Vidal, M. (2004) *Science* **303,** 540-3.

8.  Rain, J. C., Selig, L., De Reuse, H., Battaglia, V., Reverdy, C., Simon, S., Lenzen, G., Petel, F., Wojcik, J., Schachter, V., Chemama, Y., Labigne, A. & Legrain, P. (2001) *Nature* **409,** 211-5.

9.  Rual, J. F., Venkatesan, K., Hao, T., Hirozane-Kishikawa, T., Dricot, A., Li, N., Berriz, G. F., Gibbons, F. D., Dreze, M., Ayivi-Guedehoussou, N., Klitgord, N., Simon, C., Boxem, M., Milstein, S., Rosenberg, J., Goldberg, D. S., Zhang, L. V., Wong, S. L., Franklin, G., Li, S., Albala, J. S., Lim, J., Fraughton, C., Llamosas, E., Cevik, S., Bex, C., Lamesch, P., Sikorski, R. S., Vandenhaute, J., Zoghbi, H. Y., Smolyar, A., Bosak, S., Sequerra, R., Doucette-Stamm, L., Cusick, M. E., Hill, D. E., Roth, F. P. & Vidal, M. (2005) *Nature*.

10. Stelzl, U., Worm, U., Lalowski, M., Haenig, C., Brembeck, F. H., Goehler, H., Stroedicke, M., Zenkner, M., Schoenherr, A., Koeppen, S., Timm, J., Mintzlaff, S., Abraham, C., Bock, N., Kietzmann, S., Goedde, A., Toksoz, E., Droege, A.,

Krobitsch, S., Korn, B., Birchmeier, W., Lehrach, H. & Wanker, E. E. (2005) *Cell* **122,** 957-68.

11. Uetz, P., Giot, L., Cagney, G., Mansfield, T. A., Judson, R. S., Knight, J. R., Lockshon, D., Narayan, V., Srinivasan, M., Pochart, P., Qureshi-Emili, A., Li, Y., Godwin, B., Conover, D., Kalbfleisch, T., Vijayadamodar, G., Yang, M., Johnston, M., Fields, S. & Rothberg, J. M. (2000) *Nature* **403,** 623-7.

12. Barabasi, A. L. & Bonabeau, E. (2003) *Sci Am* **288,** 60-9.

13. Barabasi, A. L. & Albert, R. (1999) *Science* **286,** 509-12.

14. Barabasi, A. L. & Oltvai, Z. N. (2004) *Nat Rev Genet* **5,** 101-13.

15. Caldarelli, G., Capocci, A., De Los Rios, P. & Munoz, M. A. (2002) *Phys Rev Lett* **89,** 258702.

16. Eisenberg, E. & Levanon, E. Y. (2003) *Phys Rev Lett* **91,** 138701.

17. Qian, H. (2006) *J Math Biol* **52,** 277-89.

18. Fields, S. & Song, O. (1989) *Nature* **340,** 245-246.

19. Nelson, D. L. & Cox, M. M. (2004) *Lehninger Principles of Biochemistry* (W. H. Freeman.

20. Ghaemmaghami, S., Huh, W. K., Bower, K., Howson, R. W., Belle, A., Dephoure, N., O'Shea, E. K. & Weissman, J. S. (2003) *Nature* **425,** 737-41.

21. Dill, K. A. (1997) *J Biol Chem* **272,** 701-4.

22. Han, J. D., Dupuy, D., Bertin, N., Cusick, M. E. & Vidal, M. (2005) *Nat Biotechnol* **23,** 839-44.

23. Deeds, E. J., Ashenberg, O. & Shakhnovich, E. I. (2005) *Proc Natl Acad Sci U S A*.

24. Suthram, S., Sittler, T. & Ideker, T. (2005) *Nature* **438,** 108-12.

25. Consortium, I. H. G. S. (2001) *Nature* **409,** 860-921.

26. Dyson, H. J. & Wright, P. E. (2005) *Nat Rev Mol Cell Biol* **6,** 197-208.

27. Bomsztyk, K., Denisenko, O. & Ostrowski, J. (2004) *BioEssays* **26,** 629-638.

28. Pandey, N., Ganapathi, M., Kumar, K., Dasgupta, D., Das Sutar, S. K. & Dash, D. (2004) *Bioinformatics* **20,** 2904-10.

29. Ostrowski, J., Schullery, D. S., Denisenko, O. N., Higaki, Y., Watts, J., Aebersold, R., Stempka, L., Gschwendt, M. & Bomsztyk, K. (2000) *J.Biol.Chem.* **275,** 3619-3628.

30. Schullery, D. S., Ostrowski, J., Denisenko, O. N., Stempka, L., Shnyreva, M., Suzuki, H., Gschwendt, M. & Bomsztyk, K. (1999) *J.Bio.Chem.* **274,** 15101-15109.

31. Van Seuningen, I., Ostrowski, J., Bustelo, X., Sleath, P. & Bomsztyk, K. (1995) *J.Biol.Chem.* **270,** 26976-26985.

32. Mikula, M., Dzwonek, A., Karczmarski, J., Rubel, T., Dadlez, M., Wyrwicz, L. S., Bomsztyk, K. & Ostrowski, J. (2006) *Proteomics* **6,** 2395-406.

33. Bomsztyk, K., Van Seuningen, I., Suzuki, H., Denisenko, O. & Ostrowski, J. (1997) *FEBS Lett.* **403,** 113-115.

34. Boguna, M. & Pastor-Satorras, R. (2003) *Phys Rev E Stat Nonlin Soft Matter Phys* **68,** 036112.

35. Su, Z., Osborne, M. J., Xu, P., Xu, X., Li, Y. & Ni, F. (2005) *Biochemistry* **44,** 16461-74.

36. Liu, J. & Stormo, G. D. (2005) *BMC Bioinformatics* **6,** 176.

37. Miyazawa, S. & Jernigan, R. L. (1996) *J Mol Biol* **256,** 623-44.

38. Lee, C. L., Lin, C. T., Stell, G. & Wang, J. (2003) *Phys Rev E Stat Nonlin Soft Matter Phys* **67,** 041905.

39. Louis, M. & Becskei, A. (2002) *Sci STKE* **2002,** PE33.

40. Fiering, S., Whitelaw, E. & Martin, D. I. (2000) *Bioessays* **22,** 381-7.

41. Becskei, A., Seraphin, B. & Serrano, L. (2001) *Embo J* **20,** 2528-35.

**Figure Legends**

**Fig. 1. Strategy used to derive free energy distribution of binary protein-protein interaction from large-scale Y2H data sets.**

**Fig.2. Degree distribution of Y2H protein-protein interactions.** Number of proteins, N, with a given number of links from Y2H screen of *H. pylori* (8), the malaria parasite *P.falciparum* (6), yeast *S. cerevisiae* (5, 11), worm *C. elegans* (7), fruit fly *D. melanogaster* (4) and human (9, 10) proteins, shown as dots, was used to model using simulation (shown as open circles) and semi-analytical approaches (solid line).

**Fig. 3. Dependence of degree distributions on the parameter $\lambda$ (*A*) and $\mu$ (*B*). (*A*):** Effects of varying $\lambda$. Solid (red) $\lambda=1$, short dash (light green) $\lambda=1.5$, long dash (aqua) $\lambda=2$. The value of $\mu$ is fixed at 10. **(*B*):** The parameter $\lambda$ is held fixed at 1.0 while $\mu$ is varied between 7 (upper curve), and 10 (lower curve) in steps of unity. Solid (red) $\mu=7$, short dash (light green) $\mu=8$, medium dash (green) $\mu=9$, long dash (blue) $\mu=10$. The value $\lambda$ is fixed at 1.

**Fig. 4. Differences in the indicator of mean strength of binary protein-protein interaction $\mu$, across evolution.** $\mu$ is plotted as a function of divergence times.

**Fig. 5. Cross-species comparison of free-energy distribution of Y2H protein-protein interactions.** The analytically-derived fit of the Y2H protein-protein interaction data was used to generate association free energy distribution for each species.

**Table 1**

| Species | DT | N | $\lambda$ | $\mu$ | Chisq |
|---|---|---|---|---|---|
| *H. pylori* | 3 | 732 | 0.88 | 7.06 | 0.44 |
| *P.falciparum* | 1 | 1310 | 0.93 | 7.77 | 0.49 |
| *S. cerevisiae* | 1 | 4386 | 1.18 | 7.94 | 1.72 |
| *C. elegans* | 0.7 | 2800 | 1.29 | 8.19 | 0.61 |
| *D. melanogaster* | 0.7 | 2806 | 1.53 | 8.89 | 0.06 |
| Human (Raul et.al.) | 0.1 | 1494 | 0.64 | 10.6 | 0.72 |
| Human (Stezl et.al.) | 0.1 | 1705 | 0.67 | 10.2 | 0.60 |

DT- divergence times (billion years)
N – number of proteins

$1/\lambda$ – standard deviation from the mean of binary protein-protein interaction for a given species.

$\mu$ – reflects mean strength of binary protein-protein interaction for a given species.

**Fig. 1**

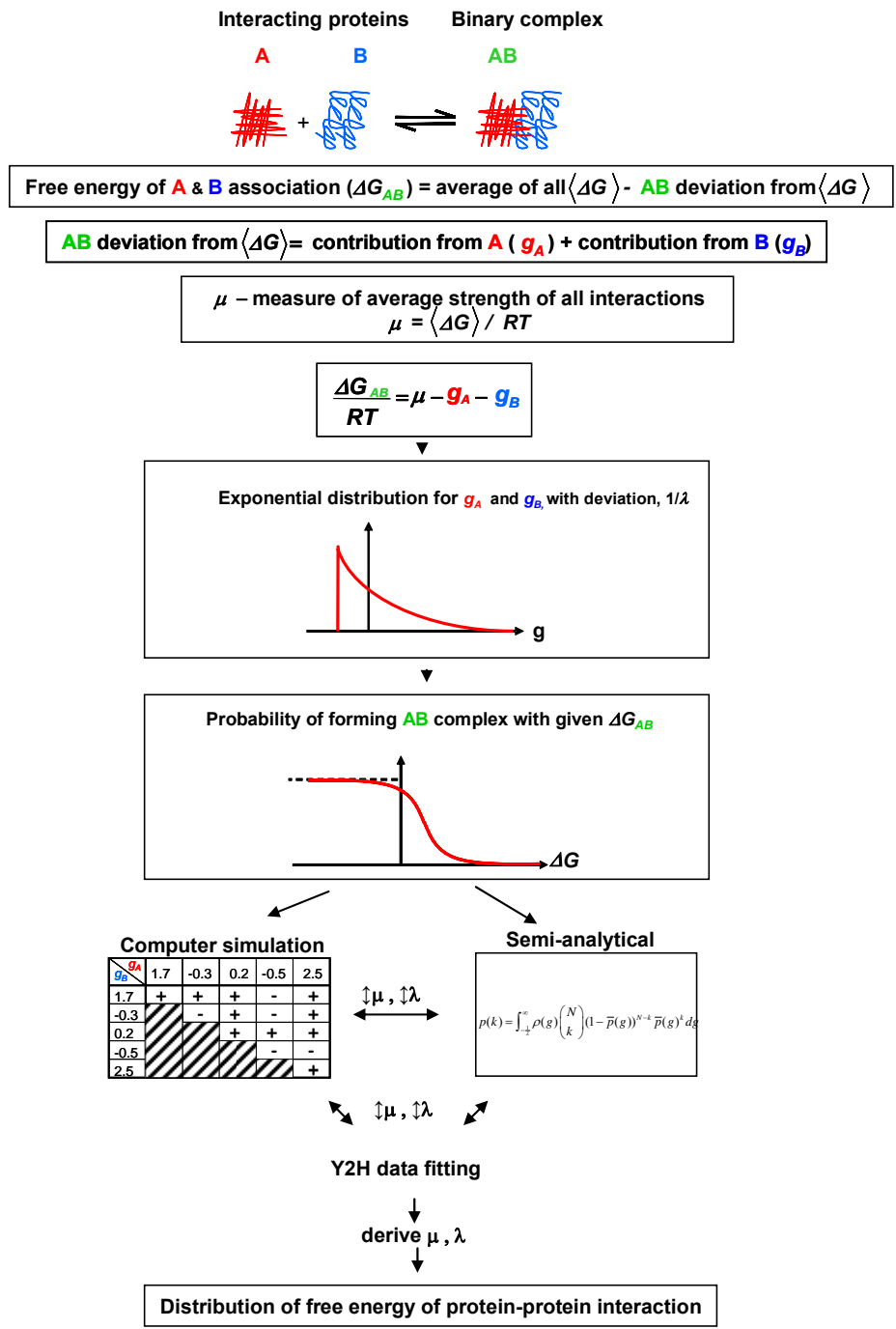

**Fig. 1. Strategy used to derive free energy distribution of binary protein-protein interaction from large-scale Y2H data sets.**

**Fig.2**

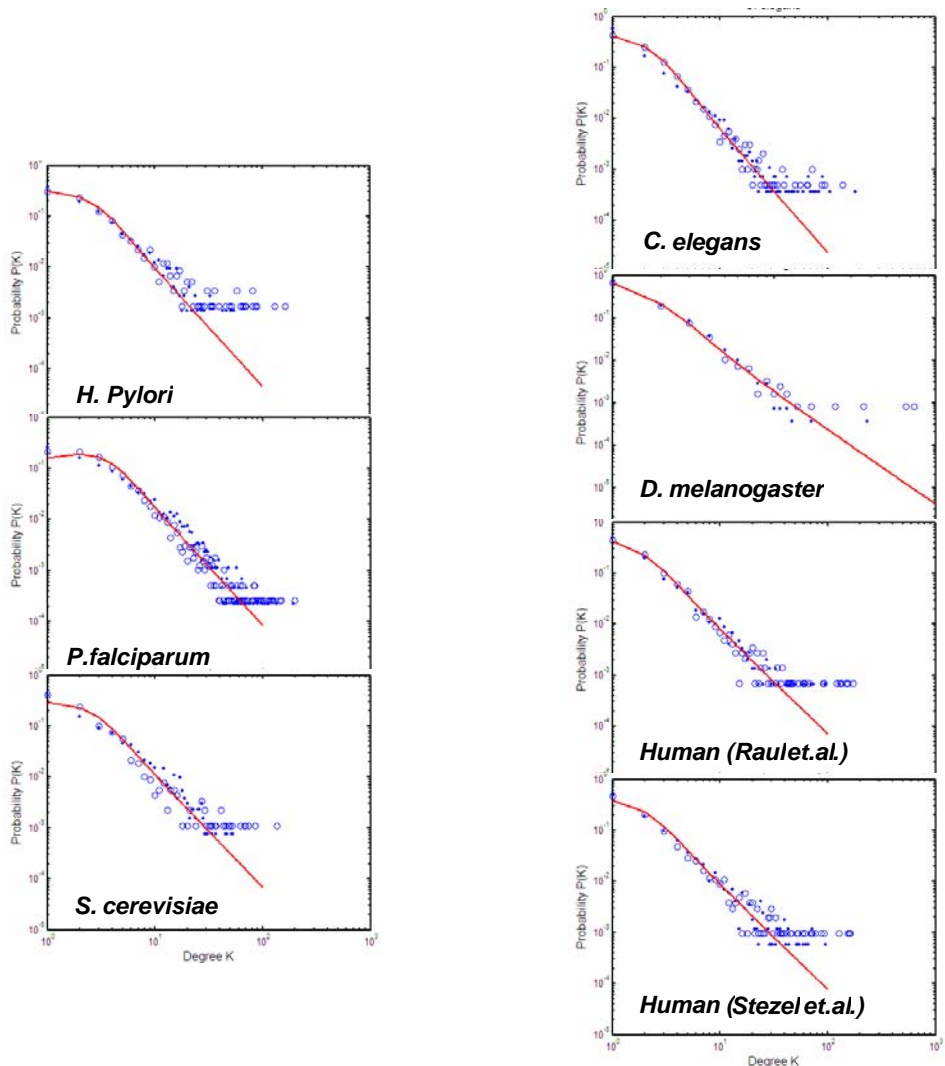

**Fig.2. Degree distribution of Y2H protein-protein interactions.** Number of proteins, N, with a given number of links from Y2H screen of *H. pylori* (8), the malaria parasite *P.falciparum* (6), yeast *S. cerevisiae* (5, 11), worm *C. elegans* (7), fruit fly *D. melanogaster* (4) and human (9, 10) proteins, shown as dots, was used to model using simulation (shown as open circles) and semi-analytical approaches (solid line).

Fig. 3

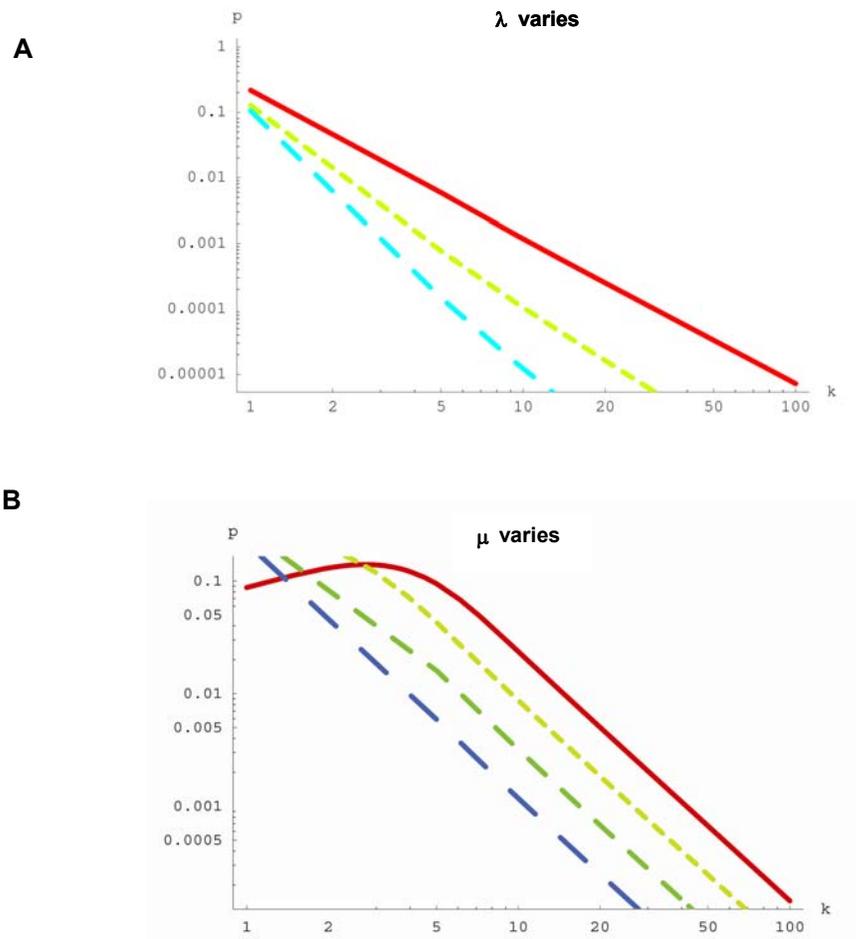

**Fig. 3. Dependence of degree distributions on the parameter λ (*A*) and μ (*B*). *A*,** Effects of varying λ. Solid (red) λ =1, short dash (light green) λ=1.5, long dash (aqua) λ.=2. The value of $\mu$ is fixed at 10. ***B*** The parameter λ is held fixed at 1.0 while μ is varied between 7 (upper curve), and 10 (lower curve) in steps of unity. Solid (red) μ = 7, short dash (light green) μ= 8, medium dash (green) μ= 9, long dash (blue) μ=10. The value λ is fixed at 1.

**Fig.4**

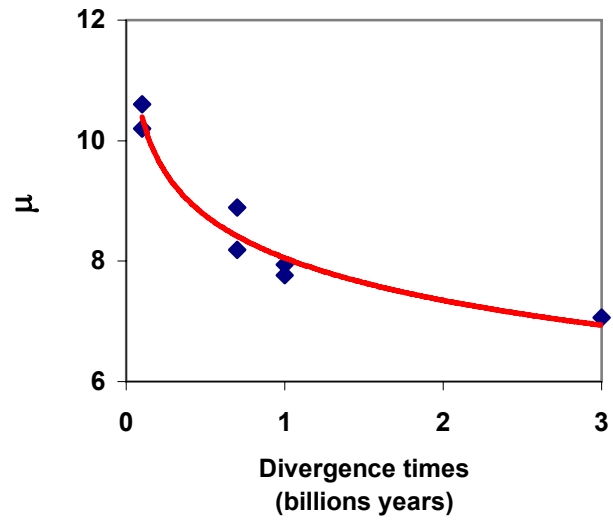

**Fig. 4. Differences in the indicator of mean strength of binary protein-protein interaction, μ, across evolution.** μ is plotted as a function of divergence times.

**Fig.5**

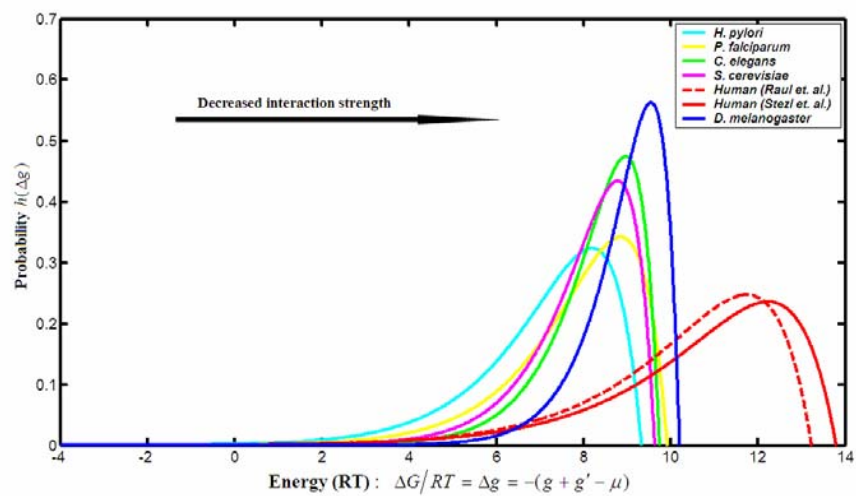

**Fig. 5. Cross-species comparison of free-energy distribution of Y2H protein-protein interactions.** The analytically-derived fit of the Y2H protein-protein interaction data was used to generate association free energy distribution for each species.